\documentclass[twocolumn,letterpaper,floatfix,showpacs,preprintnumbers,amsmath,amssymb]{revtex4}    

\usepackage[dvips]{graphicx}

\begin{document}

\def\cqg#1#2#3#4{{\it Class. Quant. Grav.} vol. #1, pp. #2-#3 (#4)}
\def\prl#1#2#3#4{{\it Phys. Rev. Lett.} vol. #1, pp. #2-#3 (#4)}
\def\prd#1#2#3#4{{\it Phys. Rev.} D vol. #1, pp. #2-#3 (#4)}
\def\apj#1#2#3#4{{\it Astrophys. J.} vol. #1, pp. #2-#3 (#4)}
\def\npb#1#2#3#4{{\it Nucl. Phys.} B vol. #1, pp. #2-#3 (#4)}
\def\ijmpc#1#2#3#4{{\it I.J.M.P.} C vol. #1, pp. #2-#3 (#4)}
\def\mnras#1#2#3#4{{\it M.N.R.A.S.} vol. #1, pp. #2-#3 (#4)}

\title{Moving black holes via singularity excision}

\author{Deirdre Shoemaker}
\affiliation{Center for Radiophysics and Space Research,
Cornell University, Ithaca NY 14853}

\author{Kenneth Smith}
\affiliation{Centers for Gravitational Physics \& Geometry and
for Gravitational Wave Physics, 
Penn State University, University Park, PA 16802}

\author{Ulrich Sperhake}
\affiliation{Centers for Gravitational Physics \& Geometry and
for Gravitational Wave Physics, 
Penn State University, University Park, PA 16802}

\author{Erik Schnetter}
\affiliation{Theoretische Astrophysik, Universit\"at T\"ubingen, 
Auf der Morgenstelle, 72076 T\"ubingen, Germany}

\author{David Fiske}
\affiliation{Department of Physics,
University of Maryland, 
College Park, MD 20742}

\author{Pablo Laguna}
\affiliation{Centers for Gravitational Physics \& Geometry and
for Gravitational Wave Physics, 
Penn State University, University Park, PA 16802}

\begin{abstract}
We present a singularity excision algorithm appropriate for numerical
simulations of black holes moving throughout the computational domain.
The method is an extension of the excision procedure previously used to
obtain stable simulations of single, non-moving black holes. The 
excision procedure also shares elements used in recent work to study the dynamics of a scalar
field in the background of a single, boosted black hole.
The robustness of our excision method is tested with
single black-hole evolutions using a coordinate system in which the
coordinate location of the black hole, and thus the excision boundary,
moves throughout the computational domain.
\end{abstract}


\maketitle

\section{Introduction}
\label{sec:intro}
Recent advances in numerical
relativity are expanding the dynamical range covered
by non-linear, three-dimensional simulations of general relativistic
systems.  However, we are still far from
having evolutions that successfully simulate a binary black-hole system  
starting from the in-spiral of the holes,
continuing through their merger and ending in the ring-down of the resulting
single black hole.
The obstacles defining this challenge are multiple
and not all understood or even known.
For example, it has become increasingly 
evident that the form of the Einstein equations implemented in 
numerical codes is one of the important aspects determining
the behavior of the simulations.
In principle, there are an infinite number of ways to recast
the Einstein equations as a Cauchy problem. 
The search for the formulation or family of 
formulations that will lead to long-lived evolutions
has, in some situations, become a tour-de-force.
In spite of the many ``3+1 flavors" 
of the Einstein equations so far introduced,
there has not been a formulation exhibiting a clear superiority.

Besides the form of the evolution equations,
there are other essential ingredients that have been demonstrated to
yield improvements in the duration of the simulations.
Gauge or coordinate conditions, methods to handle
the singularities, non-reflective boundary conditions and discretization
methods are some examples of these elements \cite{Luis_review, Thomas_review}.  

Another major focus in numerical relativity 
has been producing long-lasting, three-dimensional
evolutions of the simplest case of
a space-time containing black-hole singularities,
namely that of a single, excised black hole in which the location of the
black hole remains fixed on the computational grid \cite{Pitt_stable}. 
It is safe to say that this milestone has been achieved \cite{Alcubierre1,KST,Yo_stable}. 
One should keep in mind, however, that some of the elements used to
obtain simulations lasting forever
cannot be extended or applied to binary black-hole orbits and coalescences.
Specifically, the elements in question, as we will later
discuss in detail, are those
that hinge on keeping fixed the coordinate location of the
black-hole singularities and thus the excision boundary. 

For sufficiently large separations, even in the non-linear regime,
it should be possible to construct 
an approximate co-rotating coordinate system in which the coordinate drift of the black holes
is avoided via an appropriate set of gauge or coordinate conditions \cite{IBBH}. 
However, once a given co-rotating frame can no longer keep the location of the black holes
fixed without introducing extreme grid stretching, construction of a
new co-rotating frame will be needed. It is then likely that 
in this new co-rotating frame, the coordinate location of the black holes,
and therefore the excision regions, will be different. Consequently, 
there is a natural need to develop an excision algorithm that is able to accommodate 
motion of the black holes. This implies handling, among other things, 
grid points that emerge from inside of the excised regions and 
become part of the computational domain. 

There are examples of simulations with
moving excision already in the literature.
Using a characteristic formulation, 
long-lasting evolutions of a wobbling black hole have been obtained \cite{wobble}.
Other examples of simulations involving motion of the excision boundary 
include boosted, single black holes \cite{moving} and grazing collisions
\cite{Brandt}. Recently, Yo, Baumgarte and Shapiro \cite{Yo_moving} presented
a simple scheme for a moving excision boundary when solving the scalar field equation
in the background of a boosted black hole.

The objective of this paper is to enhance the excision algorithm
used in long-term stable simulations of single non-moving black holes to handle
the case of moving black holes. Our approach can be viewed as an extension
of the work in Ref.~\cite{Yo_moving}.
The paper is structured as follows: the formulation of the Einstein 
equations and the method used in our code, called Maya, to solve
these equations are summarized in \S~\ref{sec:BSSN}.
The crux of this paper, namely the implementation of excision, is given in \S~\ref{sec:maya_excision}.
Tests of the excision method  for a moving black hole
are presented in \S~\ref{sec:dancing}.
Finally, we discuss the ramifications of our
results and future developments in \S~\ref{sec:conclusions}.

\section{The Maya Code}
\label{sec:BSSN}
When viewed as a Cauchy problem, the Einstein equations provide a mechanism
to construct the time-history of the geometry of 
three-dimensional space-like hypersurfaces. The geometry of
each hypersurface is characterized by the intrinsic
metric $g_{ij}$ and extrinsic curvature
$K_{ij}$. The space-time foliation is glued together
with the help of the lapse function $\alpha$ and the shift vector $\beta^i$. 
In principle, the problem reduces to finding $g_{ij}$ and $K_{ij}$ for all 
the hypersurfaces by solving the Einstein equations written explicitly in terms
of $g_{ij}$ and $K_{ij}$, namely the Einstein equations in a 3+1 form. 
An example of these equations are the ADM equations \cite{ADM,York}.
However, evolutions obtained with the ADM equations have had very limited success.   
Nakamura and Shibata \cite{SN}, and later Baumgarte and Shapiro \cite{BS},
introduced a formulation (BSSN) of the Einstein equations that 
has clearly yielded remarkable improvements. 
Instead of $g_{ij}$ and $K_{ij}$ as primary variables, the BSSN formulation 
introduces new variables $\Phi,\, \hat g_{ij},\, K,\, \hat A_{ij}$
and $\widehat\Gamma^i$.  
The relationships between the BSSN and ADM variables are:
\begin{eqnarray}
\label{eq:Phi}
\Phi         &=& \frac{1}{6}\,\ln g^{1/2} \\
\label{eq:ghat}
\hat g_{ij}  &=& e^{- 4 \Phi}\, g_{ij} \\
\label{eq:K}
K            &=& g^{ij} K_{ij} \\
\label{eq:A}
\hat A_{ij}  &=& e^{- 4 \Phi}\,A_{ij} \\
\label{eq:gamma}
\widehat \Gamma^i &\equiv& \hat g^{jk} \widehat \Gamma^{i}_{jk}
        = - \partial_j \hat g^{ij}\,,
\end{eqnarray}
where
$A_{ij}  = K_{ij} - g_{ij}\,K/3$.
Above, Eqs.~(\ref{eq:Phi}) and (\ref{eq:ghat}) imply that
the conformal metric $\hat g_{ij}$ has unit determinant, which in turn
yields the second equality in Eq.~(\ref{eq:gamma}). 
In addition to the standard constraints, Hamiltonian and momentum constraints,
the BSSN equations require satisfying the conditions
$\sqrt{\hat g} = 1$ and $\hat A^i\,_i = 0$ if one chooses to
evolve all the tensor components of $\hat g_{ij}$ and $\hat A_{ij}$. 
Explicitly enforcing $\sqrt{\hat g} = 1$  is not necessary, but
insuring that $\hat A_{ij}$ remains 
trace-free is crucial in extending the life of numerical evolutions. 
In terms of the new variables (\ref{eq:Phi}-\ref{eq:gamma}), the Einstein
evolution equations take the following form:
\begin{eqnarray}
\label{eq:Phidot}
\partial_t \Phi &=& {\cal L}_{\beta}\Phi - \frac{1}{6} \alpha\, K \\
\label{eq:gdot}
\partial_t \hat g_{ij} &=& {\cal L}_{\beta}\hat g_{ij} - 2\, \alpha\, \hat A_{ij} \\
\label{eq:Kdot}
\partial_t K &=& {\cal L}_{\beta}K - \nabla_i \nabla^i \alpha +
        \alpha\,(\hat A_{ij} \hat A^{ij} + K^2/3) \\
\label{eq:Adot}
\partial_t \hat A_{ij} &=& {\cal L}_{\beta}\hat A_{ij}+ e^{- 4 \Phi} \left(
        - \nabla_i \nabla_j \alpha  +
        \alpha\, R_{ij} \right)^{TF} \nonumber\\
        &+& \alpha\, (K \hat A_{ij} - 2 \hat A_{il} \hat A^l_{~j})\\
\label{eq:gammadot}
\partial_t \widehat\Gamma^i &=& {\cal L}_{\beta} \widehat\Gamma^i
- 2\,\hat A^{ij}\partial_j\alpha \nonumber\\
&+& 2\,\alpha\widehat\Gamma^i_{jk}\hat A^{jk}
+ 12\,\alpha\hat A^{ij}\partial_j\Phi
- \frac{4}{3}\alpha\hat g^{ij}\partial_jK \nonumber\\
&+& \left[ \left( \chi+\frac{2}{3} \right) \hat\gamma^{kl}\widehat\Gamma^i_{kl}-
\chi\widehat\Gamma^i \right] \partial_j\beta^j\,.
\end{eqnarray}
The term in Eq.~(\ref{eq:Adot}) with superscript $TF$ denotes the trace-free part
of the tensor terms between brackets. Eq.~(\ref{eq:gammadot}) has been modified from the original
BSSN form ($\chi$-terms) following Ref.~\cite{Yo_stable}. 
This modification is crucial to achieve stability in
single, non-moving black-hole simulations. Finally, the first terms in the r.h.s. of 
Eqs.~(\ref{eq:Phidot}-\ref{eq:gammadot}) involving ${\cal L}_{\beta}$ denote Lie derivatives
of the scalar $K$ and tensor densities $\hat g_{ij}$ and $\hat A_{ij}$, 
as well as their extension to quantities related to
tensor densities, such as $\Phi$ and $\widehat \Gamma^i$.

We have developed a numerical code, called Maya, to solve
Eqs.~(\ref{eq:Phidot}-\ref{eq:gammadot}). The code views this
system of equations as having the following structure:
\begin{equation}
\partial_t u =  \beta^i\partial_i u + T(u,\,\partial \beta) 
                     + S(u,\,\partial u,\,\partial^2 u) \equiv \rho\, 
  \label{eq:mol}
\end{equation}
where $u = \{\Phi,\, \hat{g}_{ij},\, K,\, \hat{A}_{ij},\,\widehat \Gamma^i \}$.
Notice that the terms 
${\cal L}_{\beta}u$ have been explicitly separated into
$\beta^i\partial_i u + T(u,\,\partial \beta)$ terms.
In the interior of the computational domain, that is, excluding the
outer and excision boundaries, spatial derivatives appearing in $T$ and $S$ 
are approximated using second-order,
centered finite differencing.
The terms of the form $\beta^i \partial_i u$ 
are commonly called ``advection" terms. They are 
approximate via a second order or higher upwind 
differencing; for details, see Ref.~\cite{Kelly}.
  
At the outer boundary, the traditional approach for BSSN-based 
evolutions has been to assume that 
\begin{equation}
\label{eq:outcond}
u = u_o + \frac{w(t-r)}{r}\,,
\end{equation}
where $u_o$ represents an analytic solution. For the case of single 
black holes, this analytic solution is the exact solution to the Einstein
equations. If an exact solution is not available, the flat space-time
solution is used provided the outer boundary is sufficiently far away from
the holes. Eq.~(\ref{eq:outcond}) is then implemented in a differential form,
namely
\begin{equation}
\label{eq:radcond}
\partial_tu = -\frac{x^i}{r}\partial_i(u-u_o)
+ \frac{(u-u_o)}{r}\,.
\end{equation}
The r.h.s. of Eq.~(\ref{eq:radcond}) is discretized as in the
advection term in Eqs.~(\ref{eq:Phidot}-\ref{eq:gammadot}).
It was found \cite{Alcubierre1,Yo_stable} 
that in order to achieve stability for single, non-moving
black-hole evolutions, condition (\ref{eq:radcond}) should not
be used for the connection $\widehat\Gamma^i$. For this 
field, one needs instead to impose $\partial_t\widehat\Gamma^i=0$.

Notice that in deriving Eq.~(\ref{eq:radcond}), 
it was assumed that the analytic solution is time-independent. This will
not be the case for the moving black hole in our tests. For simplicity,
we have chosen instead to directly use the Dirichlet condition
$u = u_o$. This will turn out to be one of
the two factors limiting the duration of the simulations.
 
At the excision boundary, the discretization of spatial derivatives
also requires special treatment. This constitutes the focus of the 
paper and is addressed
in detail in the following section.

Once all spatial operators in Eq.~(\ref{eq:mol}) are discretized, 
$u$ is updated using a second-order, iterative Crank-Nicholson scheme
\cite{Teukolsky}.  Parallelization, IO and parameter manipulations in the Maya code
are handled via the
Cactus toolkit developed by the Albert Einstein 
Institute in Golm, Germany \cite{Cactus}.  

Finally, several studies have shown \cite{Kelly,Alcubierre1,nct,Yo_stable}
that stable evolutions of single black holes are not possible if both the
lapse function and shift vector are analytic functions from the exact solutions.
Instead, one needs ``driver" gauge conditions such as the
$1+\log$ condition for $\alpha$ or the $\Gamma$-driver for $\beta^i$; for
a review of gauge conditions, see Ref.~\cite{AEI_review}. Unfortunately, these driver-gauge
conditions have been designed having in mind time-independent solutions.
These gauge conditions will not be applicable for the excision tests
of a moving black hole under our consideration. We are then limited to
using the exact solution to specify $\alpha$ and $\beta^i$. 
Here resides the second factor limiting the duration of our simulations.

\section{Black Hole Excision}
\label{sec:maya_excision}
From causality considerations, one argues that it is completely unnecessary to
evolve the system of equations inside the event horizon of a black hole -- no
physically meaningful information can possibly propagate out to affect the
space-time outside the horizon.  
Doing so has the desirable effect of removing the black-hole singularity from 
the computational domain.
If one chooses to ignore, or excise, the interior of a black hole,
the immediate problem one must face 
is that of finding its event horizon. 
The event horizon, however, is a global object 
requiring a complete knowledge of the space-time, namely the
solution to the problem itself.
Unruh \cite{unruh} suggested
using instead the apparent horizon. Apparent horizons \cite{wald}
are defined in terms of quantities local-in-time. That is,
to find apparent horizons, one only
needs the spatial metric $g_{ij}$ and extrinsic curvature $K_{ij}$
of the space-like hypersurfaces in the 3+1 foliation. Thus, following
the history of apparent horizons naturally adapts to the
evolution of the system since only information at a given
instant of time is required.
The importance of apparent horizons in connection with
excision lies on the fact that an apparent horizon
will always be coincident or contained within the event
horizon. Therefore, as long as the region that is ignored or
excised is bounded by or contained within the apparent horizon, 
this region will also be interior to the event horizon.

Once the interior of an apparent horizon 
has been removed from the calculation, 
the next step to address is whether boundary conditions are required
at the excision boundary. If all the fields involved in the calculation have
at the excision boundary outgoing characteristics in the direction 
of the singularity, away from the computational domain, there is no need for
imposing boundary conditions. In this case, the fields at the excision boundary 
are evolved in the same form as the fields in the interior of the computational
domain.  

The problem in general relativity is that the only
characteristics that one can blindly assume to be outgoing
at the excision boundary are those of physical modes
since only those modes are causally constrained. For other modes, such as
gauge modes, the characteristics depend on the particular structure of
the evolution equations as well as the set of gauge conditions imposed. 
Manifestly hyperbolic formulations of the Einstein equations
facilitate determining the characteristics of all the fields,
thus allowing the identification of those fields requiring
conditions at the excision boundary. 
Here is one of the main reasons for the popularity of explicitly hyperbolic
formulations of the Einstein equations \cite{hyper_review}. 
This would suggest that numerical simulations involving excision are only
possible using a hyperbolic formulation. 
However, studies by several groups using the BSSN formulation, a formulation 
that is not explicitly hyperbolic, have shown \cite{BS,AEI_BSSN,BSSN_hyper} that 
this is not the case. BSSN-based codes,
assuming all fields having outgoing characteristics at the 
excision boundary, have produced results similar to those based on
hyperbolic formulations.

In summary, we work under the assumption that 
there is no need for boundary conditions at the excision boundary. 
The only task is
to design a discretization of the evolution equations that is appropriate
at the excision boundary where centered finite differencing is no 
longer feasible.

A possible approach to handle spatial derivatives 
at the excision boundary 
is to modify all finite-difference stencils at this
boundary so as to avoid using grid-points from within the excised region
(\textit{e.g.} one-sided differences), while still keeping the same order of
accuracy. This typically involves getting lost in a veritable jungle of logic,
especially when the excision region is that of the ``Lego sphere'',
a representation of a sphere using only the points on a fixed Cartesian grid.
Each type of corner must be treated with its own finite difference stencil. 
This approach was followed in early simulations \cite{moving,Brandt}.
To circumvent the complications of this stencil-modification method, Alcubierre
and Br\"{u}gmann \cite{Alcubierre1} 
introduced a very simple excision method which is ideally suited
for stationary space-times such as those involved with single black-hole
evolutions.  In \cite{Alcubierre1}, simulations of single black holes
that did not assume any symmetries had limited duration. 
It has been recently demonstrated 
\cite{Yo_stable}, however, that the factor restricting the duration of these simulations
was not the excision method used, but a term in the evolution equation
of the connection $\widehat \Gamma^i$.  

Anticipating that for more general cases, e.g.
binary black holes, the excision boundary and the r.h.s. in
Eqs.~(\ref{eq:Phidot}-\ref{eq:gammadot}) will have strong temporal
dependence, 
we present a method for handling the excision 
boundary that generalizes to non-stationary space-times. Our approach
extends the method used by
Yo, Baumgarte and Shapiro in their study of the dynamics of
a scalar field in the background of a single, boosted black hole \cite{Yo_moving}.
We have carried out tests to show that our method, when applied to
the case of a non-moving black hole, reproduces the observed
long-term stability observed in Ref.~\cite{Yo_stable}.

\begin{figure}[h]
\includegraphics[height=4cm]{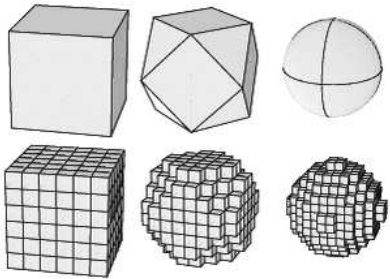}
\caption{ \label{fig:shapes}
The ``idealized'' excision shapes one may wish to use for excision - a
cube, a cuboctahedron, and a sphere, along with their representations on
a Cartesian grid. The commonly-seen ``Lego sphere'' at the bottom right
is the largest of the three shapes in terms of the volume it encloses.}
\end{figure}

Once an apparent horizon is found, the first step is
to choose the shape and size of the excision region.
A brief comment is warranted regarding the choice of the shape of the excision
region. Historically, a sphere was used. The rationale for this is that the
apparent horizon will tend to be spherical or ellipsoidal in all but extremely
dynamical space-times. As mentioned before, on the Cartesian grids typically used
in three-dimensional numerics, one achieves only a crude approximation of the sphere,
colloquially known as a ``Lego sphere''. Alcubierre and Br\"ugmann simplified
matters \cite{Alcubierre1} by using a cube as an excision shape. We have
experimented with a variety of excision shapes, all possessing at the very
least octahedral symmetry (because of the frequent use of octant, quadrant,
and equatorial symmetry boundary conditions in performing three-dimensional simulations), 
and in the end, have settled on the sphere, the cube, and the cuboctahedron.
These shapes are shown in Fig.~\ref{fig:shapes}.

\begin{figure}[h]
\includegraphics[height=6cm]{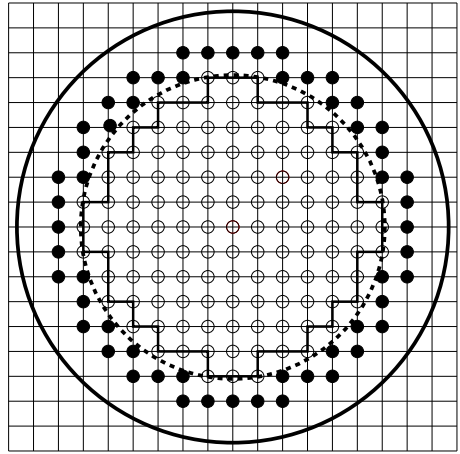}
\caption{ \label{fig:exci}
Schematic of the mask function for the case of spherical excision approximated on 
a Cartesian grid.  The large, dark circle represents the location of the apparent horizon,
while the light circle is the spherical excision region which is approximated by the
dark line.  The small, filled circles are the points on the excision boundary.
} 
\end{figure}

To facilitate the excision method, we adopt the common practice of carrying an extra
grid function, called the ``mask'', to indicate the state of a grid-point in the 
computational domain. The typical states of a grid-point are: excised, excision
boundary, interior, and outer boundary.
Fig.~\ref{fig:exci} depicts
a schematic representation of the mask variable.  The large, dark circle is the location
of the apparent horizon.  The excision region is a ``Lego sphere'' represented by the
thick line.  The filled circles are the excision boundary points, while the empty
circles are the excised points.
We also use the mask function to label the outer boundary points, but this is 
irrelevant in the present discussion on excision.

Given $u^{n}$ at the 
interior, excision boundary and outer boundary points,
the r.h.s. $\rho$ of the evolution equations (\ref{eq:mol})
for the interior grid-points are constructed 
and used to update $u^{n} \rightarrow u^{n+1}$ at these points.
Before any subsequent updating takes place, including intermediate
iterated Crank-Nicholson steps,
updating of $u^{n}$ at the excision boundary is required.

Our approach for updating data at the excision boundary
is designed to avoid 
altering the finite difference stencils used for the interior points.
This is crucial for preserving a simple code.
It consists of obtaining the data at the excision boundary via polynomial extrapolation. 
In order to simplify matters even further, we perform 
one-dimensional extrapolations along normal directions 
to the excision boundary.
Because in general the excision boundary is only a discretized 
approximation to the geometric shape desired, the normal to the excision
boundary is not aligned with the grid-points in the computational
domain. We select the direction of extrapolation to be as close as possible to the
direction of the normal for which all
data involved on the extrapolation lie on grid-points.

\begin{figure}[h]
\includegraphics[width=7cm]{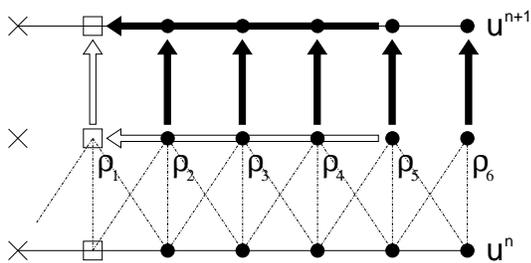}
\caption{ \label{fig:xpol}
In this schematic illustration of extrapolation onto the excision boundary,
an X represents an excised point,
a box represents an excision boundary point, and a solid dot represents
an interior point.} 
\end{figure}

We consider two approaches to construct data at the
excision boundary. One approach involves the extrapolation
of $\rho$ from the interior points to the excision boundary
(rhs-extrapolation method). The other approach is based on
extrapolation of the solution $u^{n+1}$ (sol-extrapolation method).
Fig.~\ref{fig:xpol}
shows an illustration of both methods.
Excised points are labeled with an X,
the excision boundary points with a box and interior points
with a solid dot.
As mentioned before, for each interior point 
we can use centered finite differencing
to calculate $\rho$. That is, $\rho_2$ through $\rho_6$ in 
Fig.~\ref{fig:xpol} are known,
but $\rho_1$ is unknown. The value of $\rho_1$ is, in principle, required to 
obtain $u_{1}^{n+1}$.  The rhs-extrapolation method finds the value 
$\rho_{1}$ using the values of $\rho_{2}$ through
$\rho_{5}$, depending on the order of the extrapolation.

Given that for the interior points, the truncation error 
in the discretizations to obtain $\rho$ is second-order,
the rhs-extrapolation must be such that the overall truncation error
is preserved. This implies rhs-extrapolations of at least
second-order. That is,
\begin{equation}
\rho_1 = 2\,\rho_{2} - \rho_{3} + O(h^2)\,,
\end{equation}
with $h$ the grid spacing. 
It is important to notice that the simple excision method first used 
by \cite{Alcubierre1} and later by \cite{Yo_stable}
consists of first-order extrapolation, namely
$\rho_1 = \rho_{2}$. This choice yields
truncation errors of $O(h)$. Our numerical experiments
with a moving black hole using zero-order extrapolations quickly became 
unstable, in agreement with similar results in \cite{Yo_moving}.
A possible reason why zero-order extrapolation works for non-moving
single black holes is that, from Eq.~(\ref{eq:mol}), the 
$\rho_1 = \rho_{2}$ condition is equivalent in the continuum to
\begin{equation}
0=\partial_r\rho = \partial_r\partial_t u = \partial_t\partial_r u\,,
\end{equation} 
which is consistent with the time-independent nature of the solution.  

The sol-extrapolation method uses
the values $u_2^{n+1}$ through $u_5^{n+1}$ to obtain $u_{1}^{n+1}$,
as depicted by the solid black arrow.
In order to preserve second-order accuracy of the solution in this case,
third-order extrapolations are needed. This is because the extrapolated
value $u_{1}^{n+1}$ will be used in computing 
$\rho_2$ for the updating $u_{2}^{n+1}\rightarrow u_{2}^{n+2}$.

\begin{figure}[h]
\includegraphics[height=6cm]{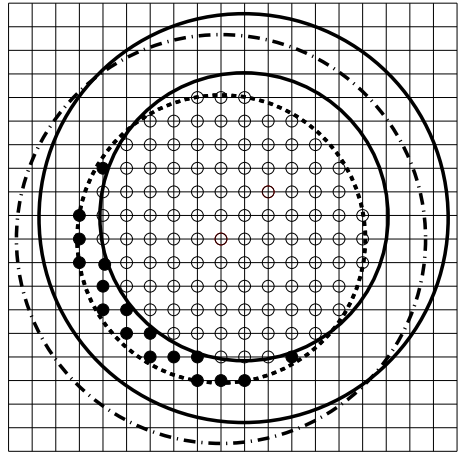}
\caption{ \label{fig:move}
This figure is similar to Fig.~\ref{fig:exci}; however, now the excision boundary has moved,
shown here by the shift in the apparent horizon.  The excision region must shift to remain
centered within the apparent horizon, thus causing a need to populate previously excised points,
represented here as filled circles.} 
\end{figure}

As the coordinate location of a black hole changes, so does
the position of the excision region. We adjust the numerical 
evolution such that the excision region does not move by more than 
a grid-point. The consequences of this change of location are that
points previously labeled excision boundary are now interior and points previously
labeled excised are now excision boundary.  Fig.~\ref{fig:move} depicts 
this process.  The movement is indicated by the shift in the apparent horizon
(dark circles), which causes the excision region to shift in order to remain centered on 
the apparent horizon.  The dotted circles are the original apparent horizon and
excision region.  The filled circles are the new excision boundary points 
that were previously excised points. 
While nothing needs to be done with interior points,
we need to populate the previously excised points with values. 
This is done following the sol-extrapolation method previously described.

\section{Dancing Black Hole}
\label{sec:dancing}
In this section, we present numerical experiments 
that demonstrate the ability of the Maya code to handle moving excision
boundaries.  Because the algorithms and data structures 
presented here are general, they can provide a preliminary treatment for 
non-trivial time-dependent systems such as black-hole binaries.

In order to test our excision algorithm, we introduce the simplest
time-dependency into the system that does not change the form
of the metric. We consider a single, non-rotating black hole
in ingoing Eddington-Finkelstein coordinates and apply a coordinate transformation
of the form:
\begin{equation}
\label{eq:coord}
x^i \rightarrow x^i + d^i\,,
\end{equation}
where 
\begin{eqnarray}
\label{eq:bouncing}
d^i &=& \left [A \sin\left(\frac{Vt}{A}\right), 0, 0 \right] \; \text{or}\\
\label{eq:circling}
d^i &=& \left [A \cos\left(\frac{Vt}{A}\right), 
A \sin\left(\frac{Vt}{A}\right), 0 \right] \,.
\end{eqnarray}
Notice that this is a coordinate transformation of the
spatial coordinates only. 
The above transformation induces the following change on the
shift vector:
\begin{equation}
\beta^i \rightarrow \beta^i + b^i\,,
\end{equation}
where 
\begin{eqnarray}
b^i &=& \left [V \cos\left(\frac{Vt}{A}\right), 0, 0 \right] \; \text{or}\\
b^i &=& \left [-V \sin\left(\frac{Vt}{A}\right), 
V \cos\left(\frac{Vt}{A}\right), 0 \right]\,.
\end{eqnarray}
The functional form of $\alpha,\, g_{ij}$ and
$K_{ij}$ remains the same. 
The only effect of this coordinate transformation is to change the
coordinate location of the black hole. For the case (\ref{eq:bouncing}),
the black hole bounces back and forth with an amplitude $A$ and 
maximum coordinate velocity $V$ (bouncing-hole). 
Similarly, for the case (\ref{eq:circling}),
the black hole moves in a circle of radius $A$ and 
coordinate velocity $V$ (circling-hole). 

The size of the excision region depends on the shape chosen.
In the case of excising a spherical region, the excision radius was set
to $1.5M$ with $M$ the mass of the black hole and $2M$ its radius,
while the cubical excision region
is inscribed to fit inside a sphere of radius $1.5M$.
That is, we are excising a larger region of the computational domain
when we use a sphere.  We have empirically found that
sol-extrapolation performs a little better than
rhs-extrapolation for the tests
under consideration. 
We use third-order ($O(h^{3})$) sol-extrapolation for the excision boundary
and fourth-order for repopulating.
The simulations were carried out with 
grid-spacings  $h=0.2M$ 
and imposing equatorial symmetry about the $z$-axis.
The computational domain 
has a total length of $20M$ in the $x$ and $y$ 
directions and $7M$ in the $z$ direction.
We used values of $A=1.0M$ and $V=0.5$ for the simulations presented here.

The driver gauge conditions commonly used to produce
long-term stability in single, non-moving black-hole evolutions
are the $1+\log$ slicing and $\Gamma$-driver; that is,
\begin{eqnarray}
\label{eq:pluslog}
\partial_t\alpha &=& \nabla_i\beta^i - \alpha\,K\\
\label{eq:gdriver}
\partial_t\beta^i &=& \lambda\,\partial_t\widehat\Gamma^i\,,
\end{eqnarray}
respectively. Unfortunately, these gauge condition do not work 
for our moving black-hole tests.
The black-hole solution that one obtains under the coordinate
transformation (\ref{eq:coord}) are not compatible with these conditions.
New or modified driver gauge conditions would have to be constructed.
These new driver conditions would likely be dependent on the particular
nature of our moving black-hole solution, thus not generalizable to 
astrophysically interesting cases such as black-hole binaries.
Because the focus of the work in this paper 
is on excision, we have chosen to set $\alpha$ and $\beta^i$ to the
analytic values given by the exact solution. It is well known
that using analytic lapse and shift gauge conditions yields
evolutions of single, non-moving
black hole lasting $<100M$. 
As we will see next,
the limits in duration of our moving black-hole simulations are consistent
with the stationary cases. 

As mentioned before, a similar problem arises
with the boundary condition (\ref{eq:radcond}). 
This condition assumes that the analytic solution is time-independent. 
For the case of time-dependent solutions, the condition becomes
\begin{equation}
\label{eq:radcond2}
\partial_tu = \partial_tu_o -\frac{x^i}{r}\partial_i(u-u_o)
+ \frac{(u-u_o)}{r}\,.
\end{equation}
The modified condition (\ref{eq:radcond2}) should in principle work for our
time-dependent case.
However, one needs to remember that the outer boundary condition
for the connection $\widehat\Gamma^i$
that yields long-term stability for non-moving holes was not 
(\ref{eq:radcond}), but $\partial_t\widehat\Gamma^i=0$.
We were unable to obtain a condition applicable to our time-dependent 
case that would lead to significant improvements in the duration of
the evolutions. Therefore,
for simplicity, we have chosen instead to directly use for all fields
the analytic solution as the boundary condition.

\begin{figure}
\includegraphics[height=8cm,width=8cm]{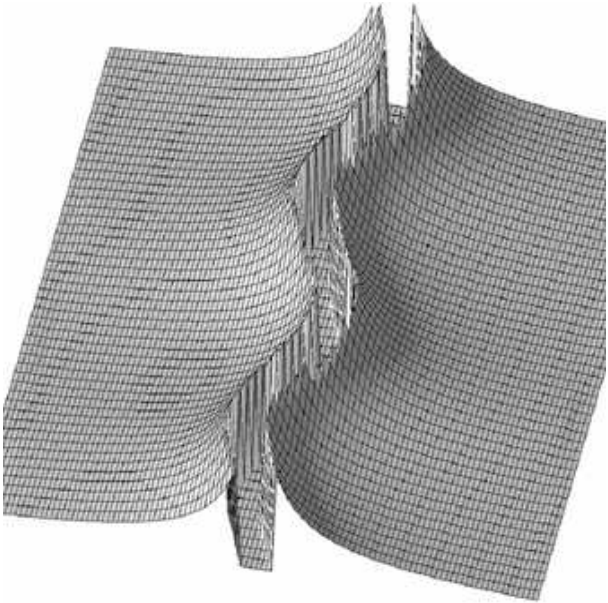}
\caption{ \label{fig:bouncing_phi}
Space-time plot of the BSSN variable $\Phi$ for a bouncing-hole.
The plot shows the evolution in the interval $0 \le t \le 25M$.
Time runs along the sinusoidal canyon.}
\end{figure}

\begin{figure}
\includegraphics[height=8cm,width=8cm]{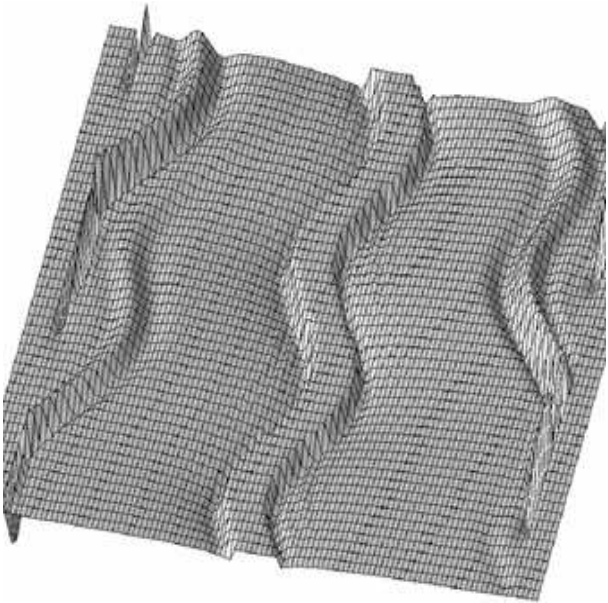}
\caption{ \label{fig:bouncing_ham}
Space-time plot of the normalized Hamiltonian constraint for  bouncing-hole.
The plot shows the evolution in the interval $0 \le t \le 25M$.
Time runs along the sinusoidal canyon.}
\end{figure}

\begin{figure}
\includegraphics[height=12cm,width=7cm]{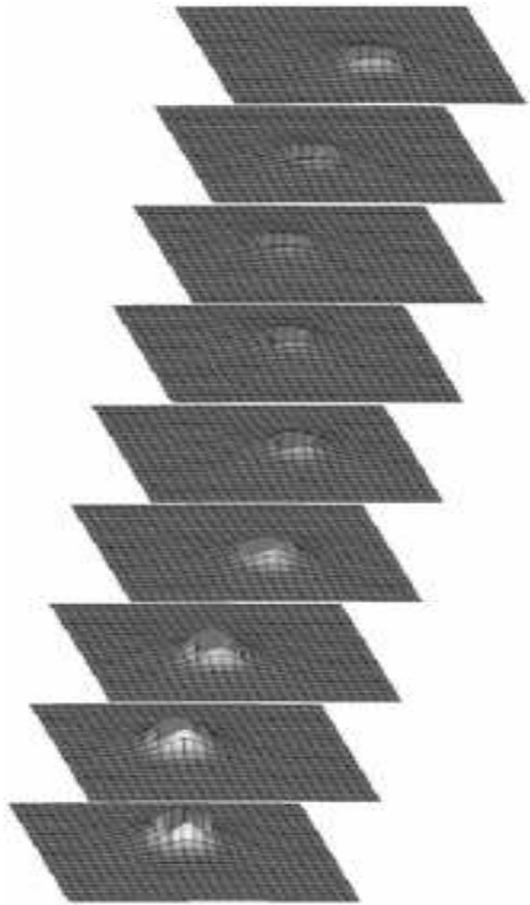}
\caption{ \label{fig:circling}
Stacked time slices of the BSSN variable $\Phi$ for a circling-hole,
with time proceeding vertically upward, and the time slices separated by $15M$.
The domain displayed only represents $\pm 5M$ in the $x$ and $y$ dimensions, for
greater clarity. The ``orbit'' of the black hole is $4\pi$, so each slice
represents an evolution of approximately one and a quarter orbits from the
slice preceding it.}
\end{figure}

Fig.~\ref{fig:bouncing_phi} is a $xt$-plot of 
the variable $\Phi$ for the bouncing-hole simulation.
The sinusoidal flat region is a $x$-cross section of the
world-tube of the excision boundary.
Fig.~\ref{fig:bouncing_ham} is also a $xt$-plot but
in this case for the normalized Hamiltonian constraint.
The reflections at the outer boundary are clear from this figure.
These effects become stronger with time and eventually
are one of the main causes of the simulation terminating.
The other contributing factors
are the analytic gauge conditions.
The circling case is depicted in Fig.~\ref{fig:circling}. 
The figure consists of stacked time slices 
of the variable $\Phi$. The time slices are separated by
$15M$ increments in time.
For clarity, the domain displayed only represents $\pm 5M$ in the $x$ and $y$ dimensions.
The ``orbit'' of the black hole is $4\pi$, so each slice
represents an evolution of approximately one and a quarter orbits from the
slice preceding it.
Animations of both bouncing and circling holes can be found at
\cite{movies}.

For comparison of evolutions, we
monitor the Hamiltonian constraint, the area of the
apparent horizon, and a ``mass" function $M_\Phi$ obtained from
\begin{equation}
\label{eq:mass}
M_\Phi = \frac{r}{2}e^{12\Phi}-1\,.
\end{equation}
For the exact solution, $M_\Phi$ is equal to the mass of the black hole.  
In all the following figures (Figs.~\ref{fig:ham_2plots}-\ref{fig:mass_phi_2plots}), 
the top panels depict results using a spherical 
excision region and the bottom cubical.  We also show results
from a stationary black-hole
case with a solid line. This is our reference case as we do 
not expect the moving cases to supersede the stationary case.
The dotted lines represent the bouncing-hole and the dashed
the circling-hole. 

\begin{figure}
\includegraphics[height=7cm,width=8cm]{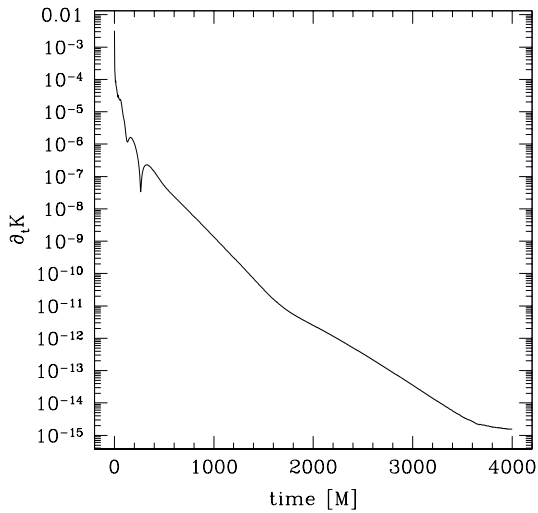}
\caption{\label{fig:stable}
L2-norm of the time derivative of $K$ for a single,
stationary black hole with driver conditions 
(\ref{eq:pluslog}-\ref{eq:gdriver})
and boundary condition (\ref{eq:radcond}).}
\end{figure}

Fig.~\ref{fig:ham_2plots} contains plots of the L2-norm of the Hamiltonian 
constraint versus time for each case.
It is evident from this figure that the simulation is unstable
even in the case of a stationary black hole.
The instability is likely due to the gauge and boundary conditions.
When we turned on the driver conditions 
(\ref{eq:pluslog}) and (\ref{eq:gdriver}),
as well as the boundary condition (\ref{eq:radcond}),
long-term stable runs for a stationary
hole were obtained, as one can see from Fig.~\ref{fig:stable},
similar to those in \cite{Yo_stable}.
Fig.~\ref{fig:ah1_area}
shows the area of the apparent horizon versus time; the apparent
horizon tracker is described in \cite{schnetter}.  The apparent horizon
tracker fails to give reasonable results at approximately $t=90M$ because
the coordinate distortions are such that the horizon intersects the excision
region at one or more points.
Finally, Fig.~\ref{fig:mass_phi_2plots} shows plots of the L2-norm
of $M_\Phi$ versus time. The reason why the L2-norm of $M_\Phi$ does not
show the jagged behavior is because the L2-norm of $M_\Phi$ was done
with data that did not include excision boundary data.
The data were taken from within a shell centered at the location of the 
black hole a distance $0.5\,M$ from the excision boundary.
The jaggedness in the Hamiltonian constraint plots is due to the 
artificial jump of values because the center of the excision
region is forced to be at a grid point, thus producing
discontinuous motion of the excision region.  

\begin{figure}
\includegraphics[height=7cm,width=8cm]{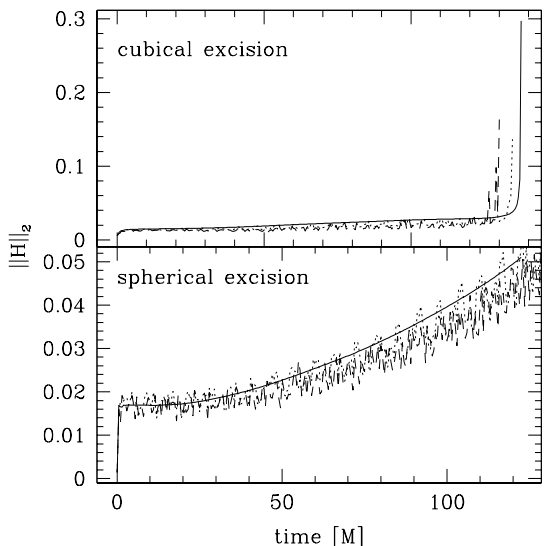}
\caption{\label{fig:ham_2plots}
L2-norm of the Hamiltonian constraint. The top panel
shows results when a cubical excision is used 
and the bottom panel corresponds to spherical excision. 
In each of the panels the solid line depicts data
from a stationary black hole, the dotted line from a bouncing
black-hole and the dashed line from a circling black-hole.}
\end{figure}
\begin{figure}
\includegraphics[height=7cm,width=8cm]{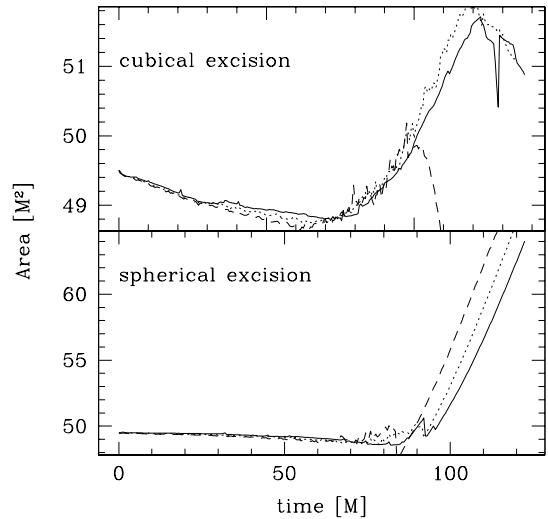}
\caption{ \label{fig:ah1_area}
Same as in Fig.~\ref{fig:ham_2plots}, but for the area of 
the black hole's apparent horizon. For reference, the
area of a single black hole is $16\,\pi\,M^2 = 50.2655\,M$.}
\end{figure}
\begin{figure}
\includegraphics[height=7cm,width=8cm]{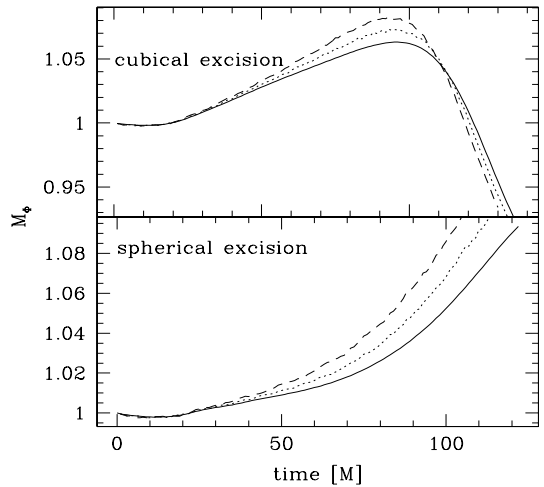}
\caption{\label{fig:mass_phi_2plots}
Same as in Fig.~\ref{fig:ham_2plots}, but for the 
L2-norm of the mass function $M_\Phi$.}
\end{figure}

\section{Conclusions}
\label{sec:conclusions}
Ref~\cite{Alcubierre1} described a simple excision algorithm that
extended the lifetimes of time-independent, single black-hole evolutions.
This work was expanded in Ref.~\cite{Yo_moving}, showing that higher order
extrapolation is necessary for stability in studying the dynamics of
a scalar field in the background of a boosted
Kerr-Schild background. 
We have generalized and applied the excision algorithm presented in \cite{Yo_moving}
to the case of a single black hole moving its coordinate location 
throughout the computational domain.  We showed
that this algorithm is generalizable to non-stationary 
black-hole space-times. 
We also find that, despite the fact that a cubical excision region
is all that is necessary for long-lived stationary black-hole runs, a spherical excision
boundary is viable.  A spherical shape may be more desirable in 
dynamic simulations because it allows the excision region to fit more easily 
within the apparent horizon.
It is important to emphasize that the algorithm in this work for excising
the black hole as it moves through the computational domain behaves 
in a similar way to excising a stationary black hole.  
This suggest that as gauge and boundary conditions suitable for
time-dependent solutions are developed, the excision method discussed in this work
can be directly implemented without requiring further or with minimal modifications. 

\section{Acknowledgments}
We thank Miguel Alcubierre, Bernd Br\"{u}gmann and Jorge Pullin 
for helpful discussions.  Parallel and I/O infrastructure provided 
by Cactus.  We acknowledge
the support of the Center for Gravitational Wave Physics funded 
by the National Science Foundation under Cooperative Agreement PHY-0114375.
Work partially supported by NSF grants PHY-9800973 to Penn State
and PHY-9800737 PHY-9900672 to Cornell University.

\end{document}